\documentclass[times,twocolumn,final,authoryear]{elsarticle}

\usepackage{jasr}
\usepackage{framed,multirow}

\usepackage{amsmath,bm}
\usepackage{amssymb}
\usepackage{latexsym}
\usepackage{soul} 

\usepackage{url}
\usepackage{xcolor}
\definecolor{newcolor}{rgb}{.8,.349,.1}
\DeclareRobustCommand{\hlcyan}[1]{{\sethlcolor{cyan}\hl{#1}}}
 
\renewcommand\underline[1]{#1} 
\renewcommand\hl[1]{#1} 

\usepackage[citebordercolor=white]{hyperref}

\journal{Advances in Space Research}

\begin{document}

\verso{N. A. Santos \textit{et. al.}}

\begin{frontmatter}

\title{First measurements of periodicities and anisotropies of cosmic ray flux observed with a
water-Cherenkov detector at the Marambio Antarctic base}%

\author[1]{Noelia Ayelén \snm{Santos}\corref{cor1}}
\ead{nsantos@at.fcen.uba.ar}
\cortext[cor1]{Corresponding author}
\author[1,2,3]{Sergio \snm{Dasso}}
\author[2,3,4]{Adriana María \snm{Gulisano}}
\author[2]{Omar \snm{Areso}}
\author[2]{Matías \snm{Pereira}}
\author[5]{Hernán \snm{Asorey}}
\author[2,6]{Lucas \snm{Rubinstein}}
\author{for the LAGO collaboration}


\address[1]{Universidad de Buenos Aires, Facultad de Ciencias Exactas y Naturales, Departamento de Ciencias de la Atm\'{o}sfera y los Oc\'{e}anos, Intendente Güiraldes 2160, C1428EGA, Ciudad Aut\'{o}noma de Buenos Aires, Argentina}

\address[2]{CONICET, Universidad de Buenos Aires, Instituto de Astronom\'{i}a y F\'{i}sica del Espacio, Intendente Güiraldes 2160, C1428EGA, Ciudad Aut\'{o}noma de Buenos Aires, Argentina}

\address[3]{Universidad de Buenos Aires, Facultad de Ciencias Exactas y Naturales, Departamento de F\'{i}sica, Intendente Güiraldes 2160, C1428EGA, Ciudad Aut\'{o}noma de Buenos Aires, Argentina}

\address[4]{Instituto Ant\'{a}rtico Argentino, Direcci\'{o}n Nacional del Ant\'{a}rtico, 25 de mayo 1143, San Martín, Buenos Aires, Argentina}

\address[5]{CONICET, Universidad Nacional de San Martin, Instituto de Tecnolog\'{i}as en Detecci\'{o}n y Astropart\'{i}culas, Centro At\'{o}mico Constituyentes, Av. Gral. Paz 1499, B1650, Villa Maipú, Buenos Aires, Argentina}

\address[6]{Universidad de Buenos Aires, Facultad de Ingenier\'{i}a, Departamento de Electr\'{o}nica, Laboratorio de Ac\'{u}stica y Electroac\'{u}stica, Av. Paseo Colón 850, C1063, Ciudad Aut\'{o}noma de Buenos Aires, Argentina}


\begin{abstract}
A new \hl{water-Cherenkov} radiation detector, located at the Argentine Marambio Antarctic Base (64.24S-56.62W), has been monitoring the variability of galactic cosmic ray (GCR) flux since 2019. \hl{ One of the main aims is to provide experimental data necessary to study interplanetary transport of GCRs during transient events at different space/time scales. } In this paper we present the detector and analyze observations made during one full year. After the analysis and correction of the GCR flux variability due to the atmospheric conditions (pressure and temperature), \hl{a study} of the periodicities is performed in order to \hlcyan{analyze} modulations due to heliospheric phenomena.
We can observe two periods: (a) 1 day, associated with the \hlcyan{Earth's} rotation combined with the spatial anisotropy of the GCR flux; and (b) $\sim$ 30 days due to solar impact of stable solar structures combined with the rotation of the Sun. From a superposed epoch analysis, and considering the geomagnetic effects, the mean diurnal  amplitude is $\sim$ 0.08\% and the maximum flux is observed in $\sim$ 15 hr local time (LT) direction in the interplanetary space.
\hl{In such a way}, we determine the capability of Neurus to observe anisotropies and other interplanetary modulations on the GCR flux \hl{arriving at the} Earth.
\end{abstract}

\end{frontmatter}


\section{Introduction}
     \label{S-Introduction} 

It is well-known that the propagation of  low-energy ($E$ $\leq$ 10 GeV) galactic cosmic rays (GCRs) is strongly influenced by the interplanetary magnetic field (IMF). Therefore, continuous monitoring of the GCR flux anisotropies and variabilities at individual ground stations is relevant
for Space Weather (SW) research \hlcyan{as it} may serve as a tool for remote sensing of the IMF ({\it e.g.} \citealp{kudela2012};  \citealp{Potgieter2013}).

There are several ground-based detector networks sensitive to different se\-condary cosmic rays (SCRs) that are generated in the particle showers initiated by primary cosmic rays (PCRs) entering the upper Earth's atmosphere. \hlcyan{For example:} Neutron Monitors (NMs), Global Muon Detector Network (GMDN), Space Environment Viewing and Analysis Network (SEVAN) and Latin America Giant \hl{Observatory} (LAGO), among others ({\it e.g.} \citealp{NM00}; \citealp{GMDN};
\citealp{LAGO17}; \citealp{SEVAN}).

The solar-controlled modulation of the
GCR flux observed at Earth is generally divided into different types according to the timescale of the variation: \hlcyan{ the 22-yr, the 11-yr, the 27-day, the diurnal variation (DV) and the Forbush-type} (\citealp{cre}). \hl{In addition to the aforementioned variations, there are several long-term periodicities} \underline{(\citealp{Potgieter2013})}\hl{ as well as a recently observed anomalous anisotropy specifically in the polar region} ({\it e.g.} \underline{\citealp{Gil2018})}.

Regarding the mean DV, it is produced by a stationary spatial anisotropy of the GCR flux in the interplanetary medium, which is observed from \hl{ground-level} stations as a \hlcyan{24-hr} temporal periodicity in SCR counting due to the \hlcyan{Earth's} rotation. Based on \hlcyan{the} co-rotation theory, which involves the equilibrium between the radial outward convection of GCRs by the solar wind and the inward diffusion along the IMF, the maximum flux is observed when looking \hlcyan{backwards} along the orbit of \hlcyan{the} Earth around the Sun, 18 hr local time (LT) direction (\citealp{parker64}; \citealp{forman}). The mean \hlcyan{DV's} amplitude observed by NMs is $\sim$0.5 \% and the maximum is in the $\sim$14-18 hr LT direction. Studies of long-term variations in the \hlcyan{DV's}  amplitude and phase suggest that it is strongly dependent on the  solar cycle stages and  the solar magnetic \hlcyan{field's} polarity ({\it e.g.} \citealp{pol}; \citealp{Anastasia}).

In addition to the solar modulation, the effects of the magnetosphere and \hlcyan{the} Earth’s atmosphere also have to be considered to make a proper interpretation of \hl{ground-level} observations. Two key concepts, such as the geomagnetic rigidity cut-off and the asymptotic cone of particle acceptance, allow us to describe all magnetospheric effects of GCRs (\citealp{nm}).

About \hlcyan{the} Earth's atmosphere, the two main causes of flux modulations are the barometric and the temperature effects. Both affect the mass distribution 
and, consequently, the associated production, absorption and decay processes of SCRs. 
The barometric effect is observed as an anticorrelation between the SCR flux and the barometric pressure variation at the observation level. This can be explained as a result of the \hlcyan{rising} absorption in the atmosphere due to the \hl{increased} mass  above the detector ($\it{e.g.}$ \citealp{PASCHALIS2013}). 

The main consequence of the temperature influence is a seasonal variation of SCR flux with a maximum in winter and a minimum in summer, particularly in ground-based detectors observing particles belonging to the electromagnetic (EM) and muonic components of the particle showers ({\it e.g.}, \citealp{Tilav}; \citealp{mendoza2016}). \hlcyan{Such muons are generated mainly by charged pions and kaons' decay}. The altitude of the muon main production layer occurs at $\approx$ 100-200 hPa (\citealp{dorman}). \hlcyan{During} summer, when the increase of this pressure level height arises due to \hlcyan{the atmosphere's expansion, the muons'} path length \hlcyan{from the generation level to the observation level} gets larger. Then, the probability of muon decay before reaching the detector increases and the observed flux decreases. The opposite occurs during winter. This is called \hlcyan{``the negative temperature effect"} and \hlcyan{it} dominates in ground-based observations.
There are different methods (empirical and theoretical) to correct SCR counting for atmospheric temperature effects, and most of them depend on observations of atmospheric variables at different altitudes (\citealp{mendoza2016}). 

In 2019, a new \hl{water-Cherenkov} detector (WCD) called Neurus was installed at the Marambio Argentinian Base in the Antarctic Peninsula for SW studies  as part of the LAGO collaboration. In this work, we empirically analyzed the pressure and temperature effects on the data recorded by Neurus. This is a crucial task for the data processing because \hlcyan{it is} only after removing \hlcyan{the} atmospheric modulations \hlcyan{that} the observed data are able to provide information on variations due to heliospheric and geomagnetic changes. In Section~\ref{S-wcd}, we describe the detector and the acquisition system. In Section~\ref{S-data}, we show the data selection. In Section~\ref{S-atm}, \hlcyan{we analyze the barometric and temperature effects using data from ERA5's atmospheric reanalysis} \underline{(\citealp{era})}. In  Section~\ref{S-mod}, we perform a spectral analysis of the corrected count rate and analyze the \hlcyan{DV's} periodicity. Finally, in Section~\ref{S-Conclusion} \hlcyan{we present the conclusions.}

\section{A water-Cherenkov Detector for SW Studies} 
      \label{S-wcd}

\subsection{SW Antarctic Laboratory at Marambio Base} 
  \label{S-lab}
A WCD called Neurus was installed at the SW Antarctic laboratory of LAMP (\hlcyan{an acronym for Spanish Laboratorio Argentino
de Meteorolog\'ia del esPacio}), which is an in\-terins\-ti\-tu\-tio\-nal group from Buenos Aires, Argentina, mainly dedicated to SW research, instrumental
development, R2O-O2R, and real\--time
monitoring of SW conditions (\citealp{20lanabere}; \citealp{lanabere21}). \hl{The scientific aims of the Antarctic SW laboratory can be found in }\citet{gulisano21}.

The laboratory was deployed during the \hlcyan{Argentinian Antarctic} campaign in the southern hemispheric summer of 2019-2020 at the Argentine Marambio Base, located at 64.24S-56.62W and 196 m \hl{above sea level}. 
Besides the particle detector, the laboratory has a magnetometer prototype, a GPS receiver to make the time stamp of observations, a meteorological station, and a telemetry system which provides 5-minutes real-time monito\-ring \hlcyan{to the servers of the group in Buenos Aires.}

Neurus WCD is part of the LAGO network (\url{lagoproject.net};
 \citealp{SIDEL}; \citealp{Dasso16}; \citealp{LAGO17}).
LAGO is a spin-off of the Pierre Auger Observatory with the concept of developing Cherenkov detectors, consis\-ting of decentralized nodes spanning over Latin America, a region of the world not covered by the majority of the ground-based detector networks. WCDs ope\-ra\-ting in counting mode are highly sensitive to Forbush Decreases (\citealp{Auger11}; \citealp{Dasso12}). They are robust, low-cost \hlcyan{and, last but not least, easy-to-maintain} and eco-friendly.

The effective geomagnetic cut-off rigidity of Marambio is $P_{c}$ = 2.1 GV (\url{https://tools.izmiran.ru/})
\hl{High-latitude stations are privileged places because they are sensitive to lower-energy GCRs and have a narrower asymptotic cone of acceptance compared with sites located at middle and low magnetic latitudes}  ({\it e.g.}\underline{\citealp{Bieber};\citealp{Mishev2020}}). \hl{Thus, they are interesting locations to study GCR  anisotropies originated during their transport  in the solar wind. Neurus, which is in a middle magnetic latitude site, it is the highest-latitude site of the LAGO collaboration.}
\hl{Besides that, it will be useful 
in the light of the recent similar measurements in the same geographic area} \underline{(\citealp{ZANINI}; \citealp{BLANCO})}.

\hl{Preliminary results from the LAGO Space Weather simulation chain can be found in} \underline{\citet{ASorey2018}} and \underline{\citet{Sarmiento2019}}. 
\hl{The authors simulate the geomagnetic effects on PCRs by using the MAGNETOCOSMICS code, the development of the air showers in the atmosphere is implemented by using the CORSIKA code, and finally a GEANT4 model is used for the detector's response. In particular, preliminary simulations for Marambio can be seen in} \underline{\citet{Sarmiento2020}}.

\hl{The count rate of background low energy particles detected by the WCDs of the  Pierre Auger observatory ($P_{c}$ $\sim$ 9.5 GV) are generated by PCRs with a median energy of about 90 GeV }(\underline{\citealp{auger15};\citealp{Dasso12}}). \hl{Taking into account that the effective magnetic rigidity cut-off for Marambio is smaller ($P_{c}$ $\sim$ 2 GV), then  the typical primary energies observed by Neurus are expected to be a few tens of GeV.}

\subsection{Neurus WCD} 
  \label{S-new}
  
\begin{figure}[ht!]
    \centering
    \includegraphics[width=8cm, height=4cm]{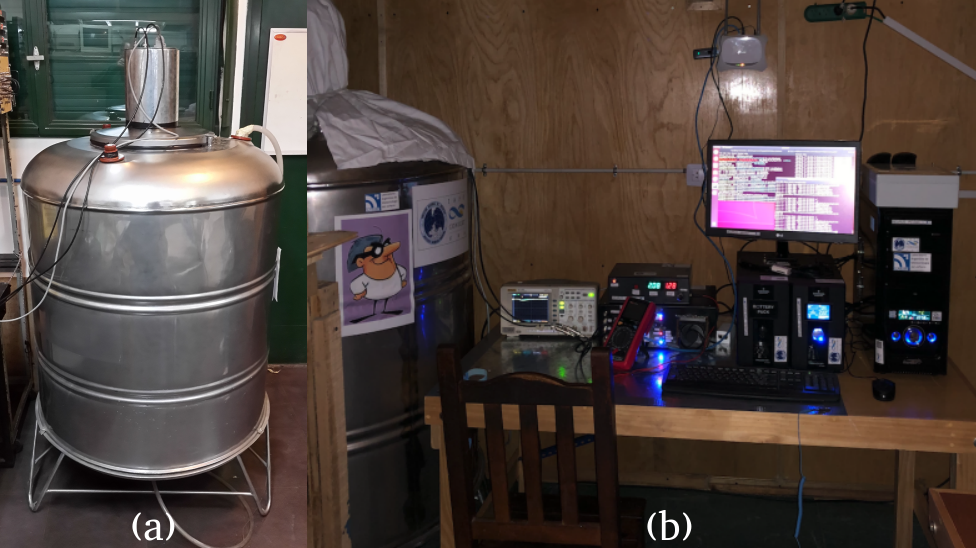}
    \caption {(a) Neurus tank developed at IAFE in Buenos Aires; (b) Same tank and the DAQ system installed at the SW Antarctic Laboratory in the Antarctic Peninsula.}
    \label{fig:fig1}
\end{figure}

The WCD is made up of three main elements: \hl{the cylindrical tank}, the PMT and the data acquisition system (DAQ) (see Figure \ref{fig:fig1}). \hl{The tank is made of stainless steel (diameter = 0.96 m, height = 1 m) and it is filled with purified water}.

\hl{This type of device is based on the possibility of detecting Cherenkov radiation from a charged particle passing through a volume of water. Charged particles from the shower of SCRs must have a minimum value of energy $E_{min}$ to radiate Cherenkov light in water. For instance, for electrons and positrons ($e^{\pm}$) and muons ($\mu^{\pm}$), which are the dominant components of the air showers at ground-level, $E_{min}^e$ $\simeq$ 0.8 MeV and $E_{min}^{\mu}$ $\simeq$ 160 MeV }(\underline{\citealp{ASorey2018}}).\hl{ Also, high-energy photons ($E_{min}^\gamma$ $\simeq$ 0.4 MeV) could be detected due to the pair production process. In recent works, it has been shown that, by doping the water, the WCD shows an enhanced detection of high-energy neutrons }(\underline{\citealp{sidelnik2020}}.)

\hl{The signal generated by muons is very different from the signal generated by electromagnetic particles. The typical energy of muons is E$_{\mu}$ $\sim$ 1 GeV and in a wide range of energies the stopping power of muons in water is $\sim$ 2 MeV cm$^{-1}$} (\underline{\citealp{auger1996}}). \hl{Therefore, the majority of muons are able to pass through the entire detector losing only a small fraction of the initial energy and, in that case, the signal produced depends only on the length that the muons traveled in water. In the case of electrons, for the typical energy E$_{e}$ $\sim$ 20 MeV, the stopping power in water is $\sim$ 2 MeV cm$^{-1}$. 
Then, the water volume is sufficient to absorb the majority of the electromagnetic particles} (\underline{\citealp{auger1996};\citealp{asoreyphd}}). 

\hl{The Cherenkov radiation produced is reflected and diffused by an internal coating made of Tyvek\textregistered. In this way, after a few reflections, all correlations between the propagation direction of the Cherenkov photons and the direction from which the particle enters to the detector is lost. This light is partially collected and amplified by a photo-multiplier tube (PMT), and then the DAQ acquires and digitizes the signal.}

Three DAQ systems were implemented: (1) \hl{an oscilloscope (Rigol DS1102E) in rate mode counts the pulses that exceed a peak threshold of 0.02 V, and then they are recorded by a communication system in a computer (Mode A), (2) a commercial board  Red Pitaya STEMLab 125-14 (14 bits resolution, 125MSPS, dynamic range $\pm$ 1V), operating as an oscilloscope, records the trace of five sample pulses per second, limited by the acquisition speed of Red Pitaya in this mode (Mode B), (3) another Red Pitaya but working as an FPGA programmed as the LAGO's DAQ system which records the trace of all pulses detected (Mode C)} (\citealp{pitaya}).

Each voltage pulse has three main characteristics which \hlcyan{result} from the convolution of the photon production Cherenkov and the response of the tank-Tyvek-PMT\hlcyan{-DAQ} system: duration (the average duration is approximately 100 ns), peak, and area under the curve (or charge). We consider 12 $\times$ $10^{7}$ vol\-tage pulses \hl{acquired} by using Mode B during April-December, 2019.

\begin{figure}[ht!]
    \centering
    \includegraphics[width=7.5cm, height=5.5cm]{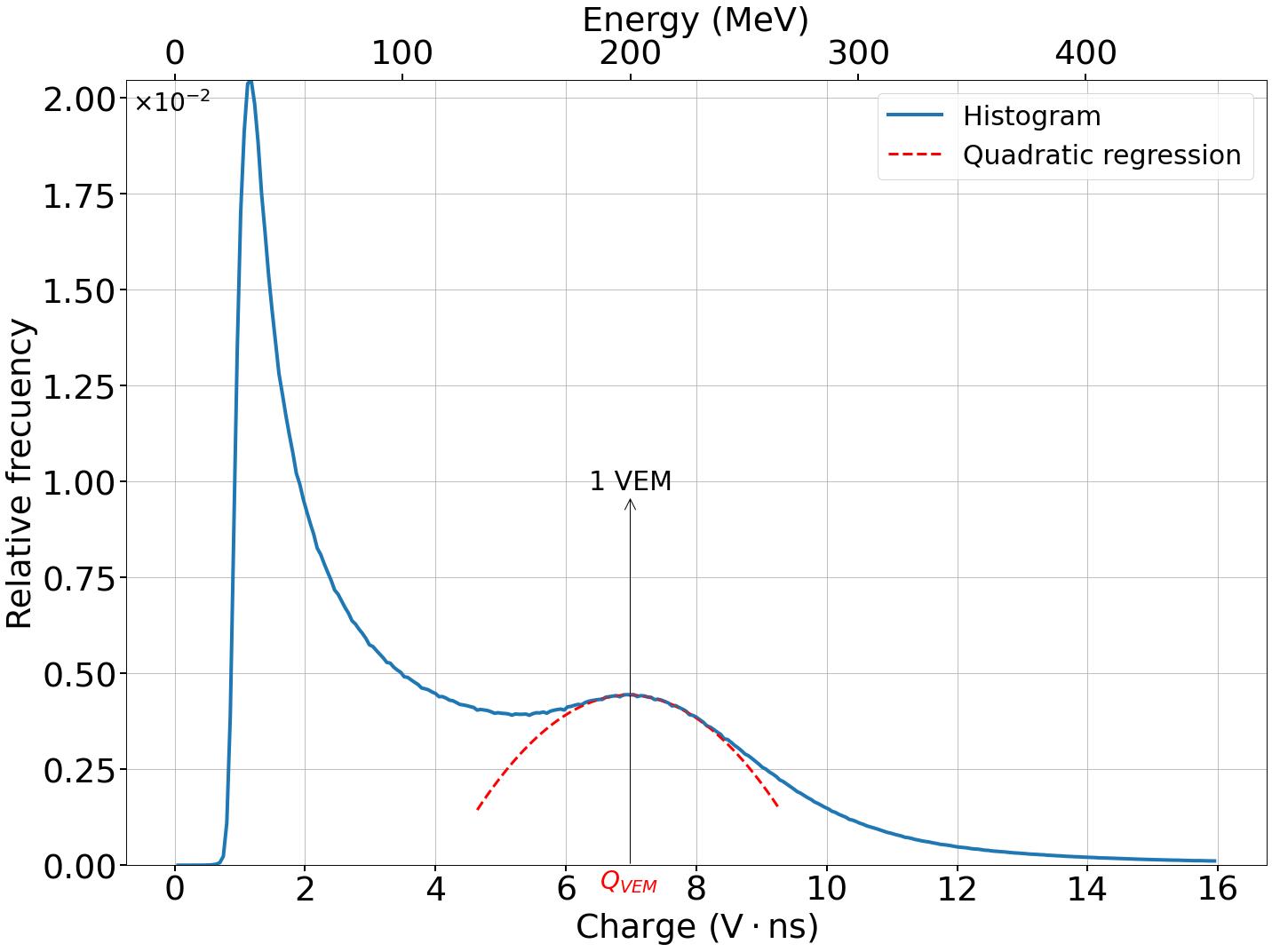}
    \caption {Normalized charge histogram for the period April-December, 2019. The  energy calibration was done by estimating the position of the second maximum (from the quadratic fit in dashed line), which is associated with the passage of vertical and central muons through the detector ($\sim$ 200 MeV).}
    \label{fig:fig2}
\end{figure}

In Figure \ref{fig:fig2}, the charge histogram is presented. This shape is the typical one observed from a WCD ({\it e.g.} \citealp{Bertou05}) and it is \hlcyan{also} well reproduced by simulations ({\it e.g.} \underline{\citealp{Sarmiento2019}}.

The first local maximum (at lower charge) is related to the chosen trigger threshold, and it is mainly originated by EM component of the showers. Values around the second local maximum 
are related to muons. Particularly, pulses associated with this maximum, called muon hump, are \hlcyan{linked} with the passage of  vertical and central muons across the detector. 
Signals \hlcyan{with} charges significantly greater than the one of the muon hump are originated by the entry of multiple particles into the detector ({\it e.g.} \citealp{asorey15}).

Since a muon in water deposits \hl{$\sim$ 2 MeV cm$^{-1}$} and conside\-ring that the water level in the tank is 1 m, it is possible to assign an energy value of \hl{$\sim$ 200 MeV} associated with pulses charge equal to the second maximum of the histogram. In \hlcyan{Figure} \ref{fig:fig2}, the position of the maximum was estimated by performing a quadratic regression  getting $Q_{VEM}$ $\approx$ 7 V $\times$ ns. The main goal of this calibration is to  obtain the value of the Vertical-Equivalent Muon (VEM) in electronics units in order to assess the quality of the measurements \hlcyan{and also to compare them} with Monte Carlo simulations.

\section{Data Selection}
\label{S-data}

In this work we use the hourly-average total count rate ($S$) measured by Neurus at Marambio, the hourly surface pressure measurement ($P$), and the room temperature ($RT$). $S$ is built from the data acquired by the oscilloscope mode, which is the only DAQ system that has been working since the detector began its observations. We also use \hlcyan{ERA5's} atmospheric reanalysis for the air column above Marambio.

In Figure \ref{fig:fig3}, we show the observations from the installation (April 2019) to March 2021. We divided the whole period into three sub-periods to highlight and summarize the DAQ systems, the set-up of the observatory, and the improvements made in each one, namely:

\begin{figure}[ht!]
    \centering
    \includegraphics[width=7cm, height=11cm]{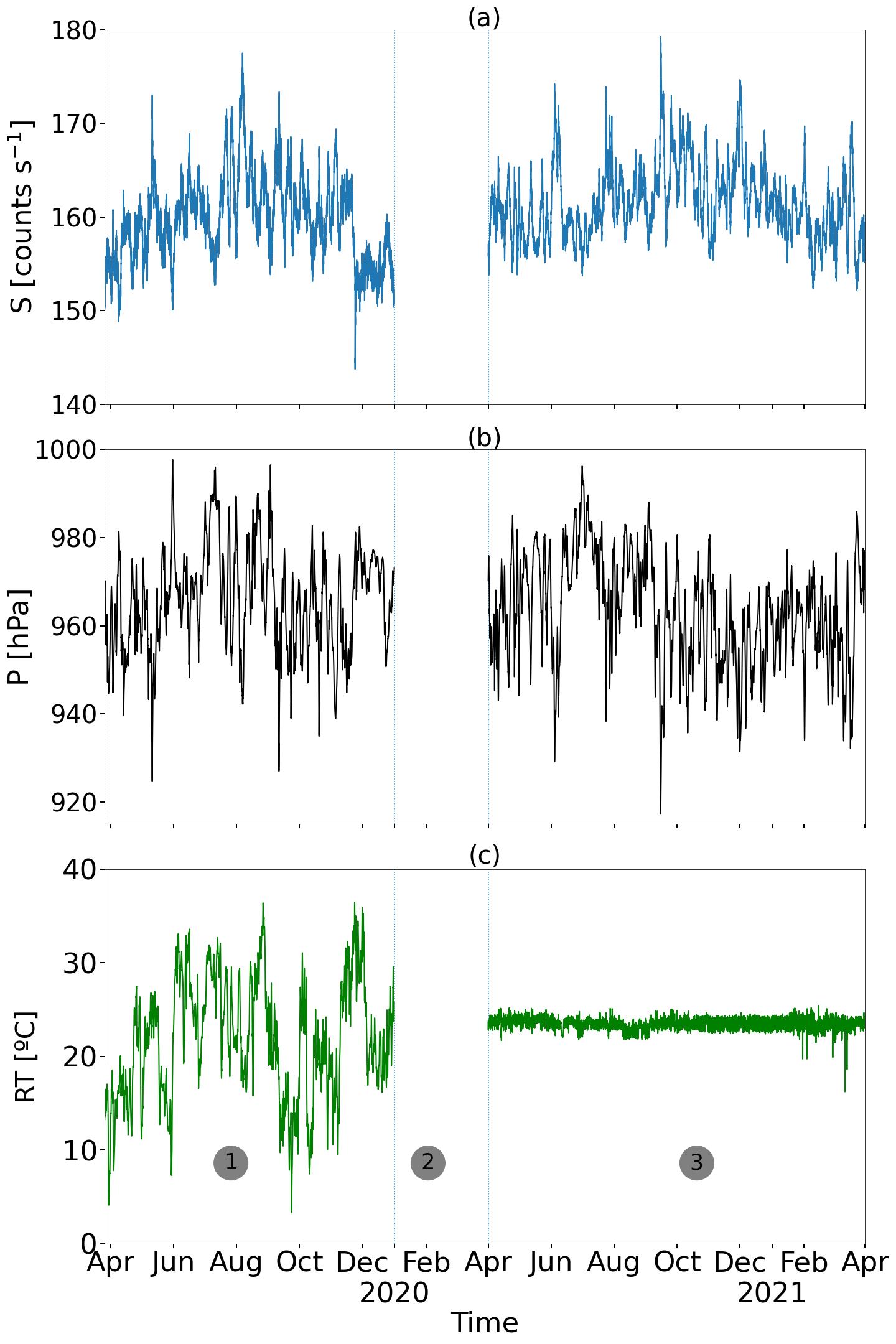}
    \caption{Hourly-average raw total count rate $S$ (a), surface pressure measurement $P$ (b), and laboratory room
temperature $RT$ (c) from April 2019 to March 2021. The data are divided in three periods depending on the lab conditions. In this work data from the third period are used. }
    \label{fig:fig3}
\end{figure}

\hlcyan{(1): From April 1, 2019, to December 31, 2019.} Two of the systems described in the previous section worked during this period: \hl{Mode A and B.} There was no laboratory room temperature control.

\hlcyan{(2): From January 1, 2020, to March 31, 2020 (2020 Antarctic campaign)}. The acquisition was stopped to carry out the updates.

\hlcyan{(3): From April 1, 2020, to March 31, 2021}. \hl{The Mode C and a laboratory room temperature control system were installed.
The second update was carried out during the 2021 Antarctic campaign without stopping the acquisition. All the described modes (A, B and C) have been operational since then.}

In this work, we analyze the data of the third period, in which the room temperature was stable to avoid different quali\-ties of the measurements due to changes in the electronic devices' behaviour.
In this period, the mean values of these three variables are: 161.5 counts s$^{-1}$, 963 hPa and 25.6$^\circ$ C for $S$, $P$, and $RT$, respectively. 

\section{Analysis of the Atmospheric Effects on Counting Particles}
\label{S-atm}

\subsection{Barometric Effect}
\label{bar}
The particle counting $S$ of any SCR component varies with a
small change in the atmospheric pressure $P$ as $dS = \mu dP$,
\hl{where $\mu$ is a negative coefficient that represents the absorption of the secondary component under consideration.} 
\hlcyan{By integrating this expression, supposing  that for pressure $P_0$ the measured intensity is $S_0$, and assuming $\mu$  = const, an exponential dependence is obtained, as seen in Equation} (\ref{eq:eq1}) (\citealp{cre}).

\begin{equation}
    S(t)=S_0 e^{ \mu (P(t) - P_0)}
    \label{eq:eq1}
\end{equation}    


Considering the first order approximation, Equation (\ref{eq:eq1}) could be expressed as:

\begin{equation}
    \frac{\Delta S}{S_{0}} \times 100 \%  = \frac{S(t) - S_0}{S_0} \times 100 \% \simeq \beta (P(t) - P_0) = \beta \Delta P 
    \label{eq:eq2}
\end{equation}    

$\Delta S /S_0$ is the count rate relative variation and $\beta$ is the barometric coefficient expressed in $\%$ hPa$^{-1}$. 

We analyze the barometric effect during the time period April 2020-March 2021 splitting it into monthly intervals, $P_0$, $S_0$ being the mean values of $P$, $S$ for each month, respectively.  

\begin{figure}[!ht]
    \centering    \includegraphics[width=7cm, height=5cm]{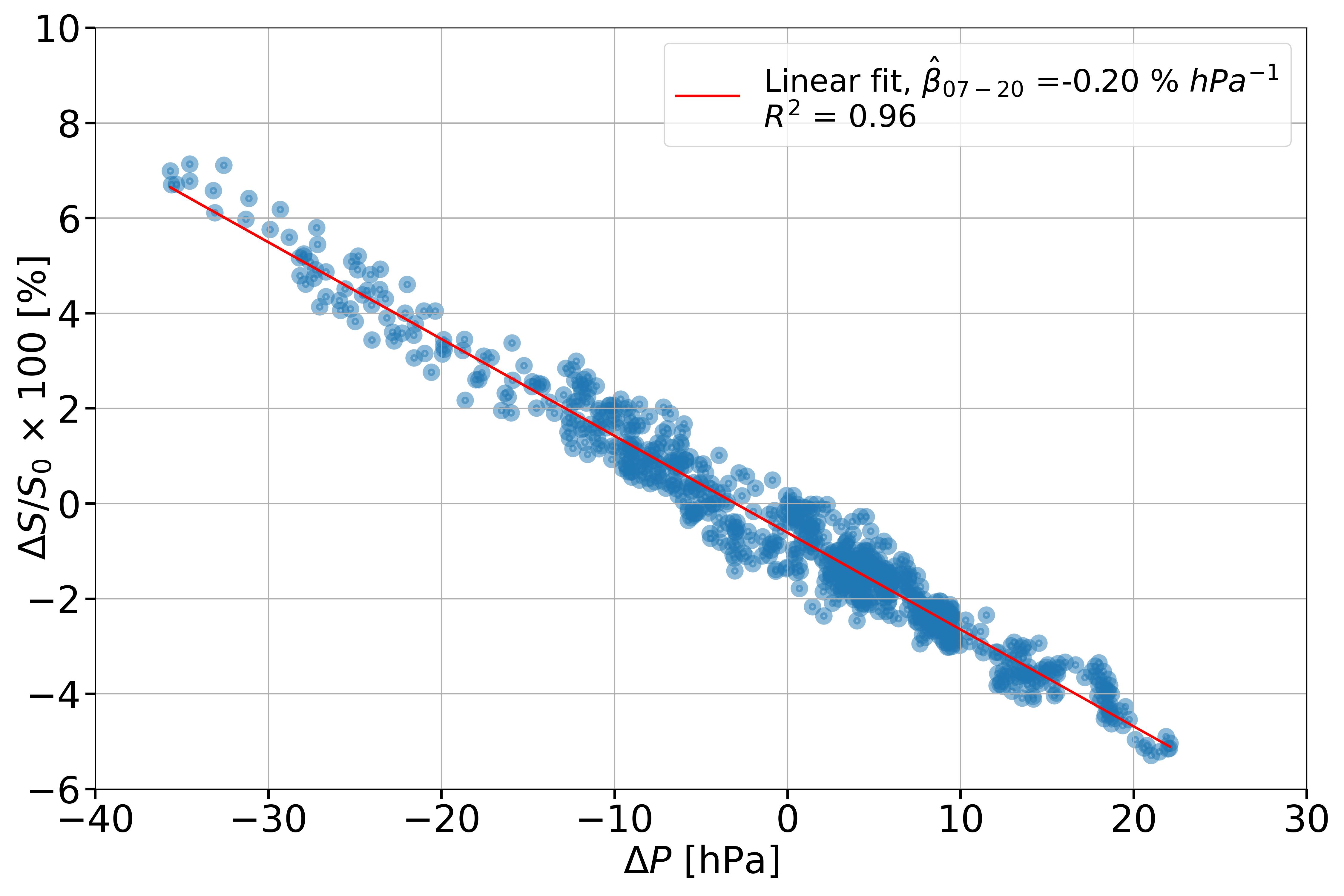}
    \caption{$\Delta S /S_0$ $\times$ 100 \% vs. $\Delta P$ for July 2020. \hlcyan{The blue points are the data and the red line is a linear fit.}}
    \label{fig:fig4}
\end{figure}

On Figure \ref{fig:fig4} there is an example of the observed data $\Delta S /S_0$ $\times$ 100 \% vs. $\Delta P$ for July 2020. \hlcyan{The expected negative correlation between both variables can be seen}, with a relative variation of the counting up to 7\% and a high Pearson correlation coefficient ($r$ = \textminus 0.9). The red line is a linear fit based on Equation(\ref{eq:eq2}) and the slope $\hat{\beta}$ corresponds to the estimated barometric coefficient. \hlcyan{Also, a linear fit was done (not shown here)} considering
Equation (\ref{eq:eq1}) and the relative difference between both $\hat{\beta}$ estimations is less than 1\% in every month, \hlcyan{then the linear approximation is good enough.}

\begin{figure}[!ht]
    \centering
    \includegraphics[width=7cm, height=10cm]{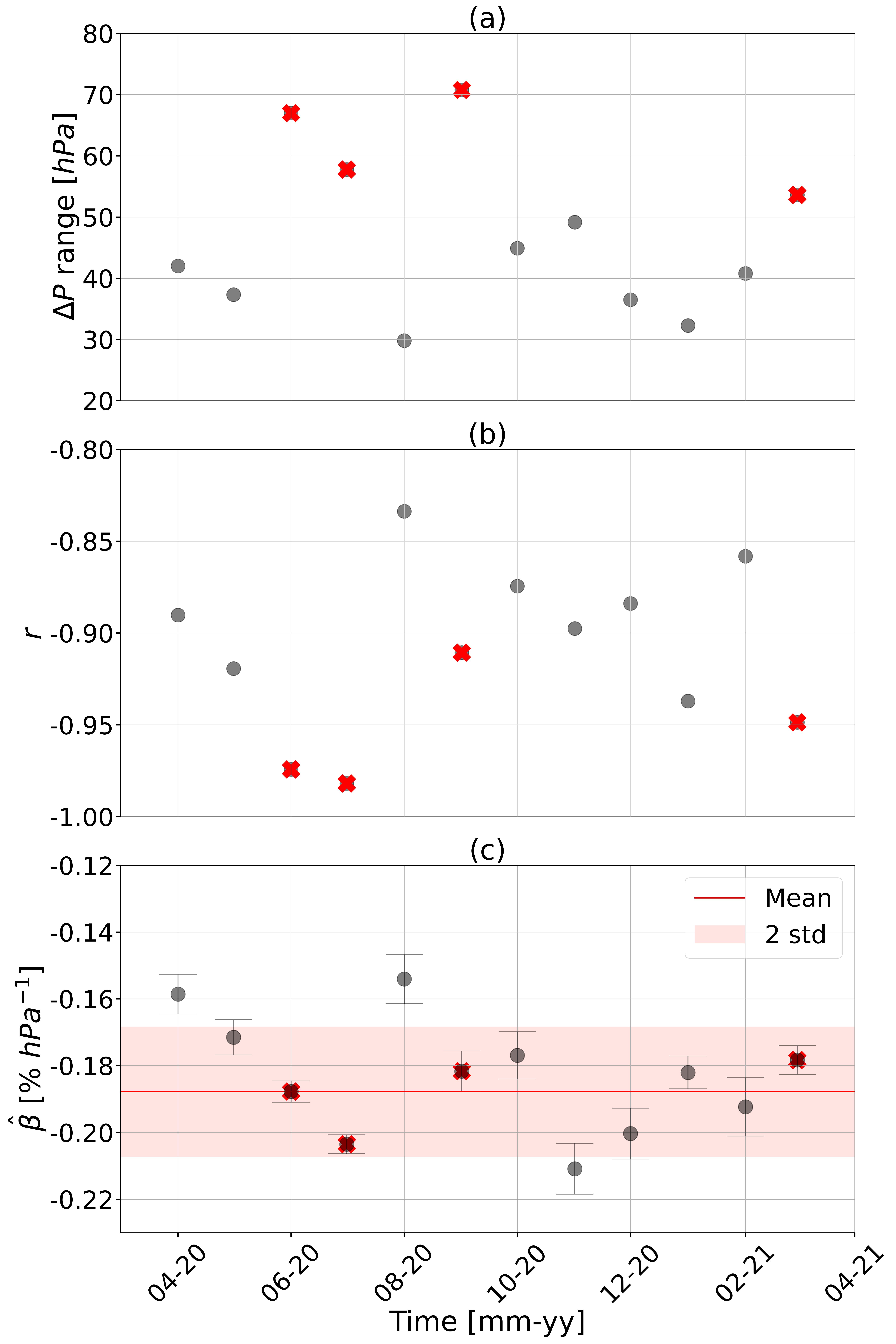}
    \caption{$\Delta P$ range (a), correlation coefficient $r$ (b), and $\hat{\beta}$ as a function of months. The red Xs are months that satisfy the chosen criteria. In (c) the mean beta value and 2 standard deviations are \hlcyan{shown.}  }
    \label{fig:fig5}
\end{figure}

On Figure \ref{fig:fig5} \hlcyan{we can see, for each month}:
($a$) pressure variation $\Delta P$ range in which we observe that the greatest variations occur during winter (big atmosphere pressure variations du\-ring the crossing of
meteorological fronts); 
($b$) the Pearson correlation coefficient $r$ between $\Delta S /S_0$ and $\Delta P$; 
($c$) the slope $\hat{\beta}$ and its 95 \% confidence interval. We estimated $<$ $\hat{\beta}$ $>$ avera\-ging monthly values and considering two standard deviation we get
$<$ $\hat{\beta}$ $>$ =   (\textminus 0.19 $\pm$ 0.02)  \% hPa$^{- 1}$.
\hlcyan{For this average, we consi\-dered} months for which  $\Delta P$ range is higher than 50 hPa. \hlcyan{This way,} these months had a considerable pressure varia\-tion and most of the variability was associated only with the barometric \-effect and not with other sources.
The four selected months are indicated with \hlcyan{a red X. We removed} the pressure \-effect from the whole time period using this coefficient $<$ $\hat{\beta}$ $>$ and we got $S'$ (counting corrected by pressure effect) using Equation (\ref{eq:eq3}).

\begin{equation}
   \frac{\Delta S'}{S_{0}} \times 100 \%  =    \frac{\Delta S}{S_{0}} \times 100 \%  - <\hat{\beta}>\Delta P
\label{eq:eq3}
\end{equation}

 $S_0$ = 161.5 counts s$^{-1}$ and $P_0$ =  963 hPa.
This barometric coefficient is consistent with other detector reports (see \citealp{Mag}; \citealp{PASCHALIS2013};
\citealp{zaz};
\citealp{mendoza2016}). 
Firstly, the barometric coefficient is much higher for the case of neutrons ($\beta$ $\sim$ \textminus 0.7 \% hPa$^{-1}$)
than the ionized components. Secondly, for these last ones $\beta$ $\sim$ \textminus (0.1–0.2) \% hPa$^{-1}$. Furthermore, the estimated value is consistent with the preliminary values estimated by MITO data, a muon telescope installed near Marambio (\citealp{mito}). This coefficient will depend on the type of particles, their energy and the zenith angle (\citealp{mendoza2019}). Thus, \hlcyan{the reported value is a global average without distinction} between any of the variables mentioned above. In future works, we will analyze the dependence of the barometric coefficient with deposited energy using the charge histograms explained in Section~\ref{S-new}.

\subsection{Temperature Effect}
\label{S-temp}

\hlcyan{The} pressure corrected data show a seasonal modulation with a maximum during August-September (winter-spring) and a minimum during February-March (summer-autumn) with an amplitude of $\sim$ 3$\%$ (see the blue curve on  Figure \ref{fig:fig7} (b)).

\begin{figure}[ht!]
    \centering
    \includegraphics[width=7cm, height=9cm]{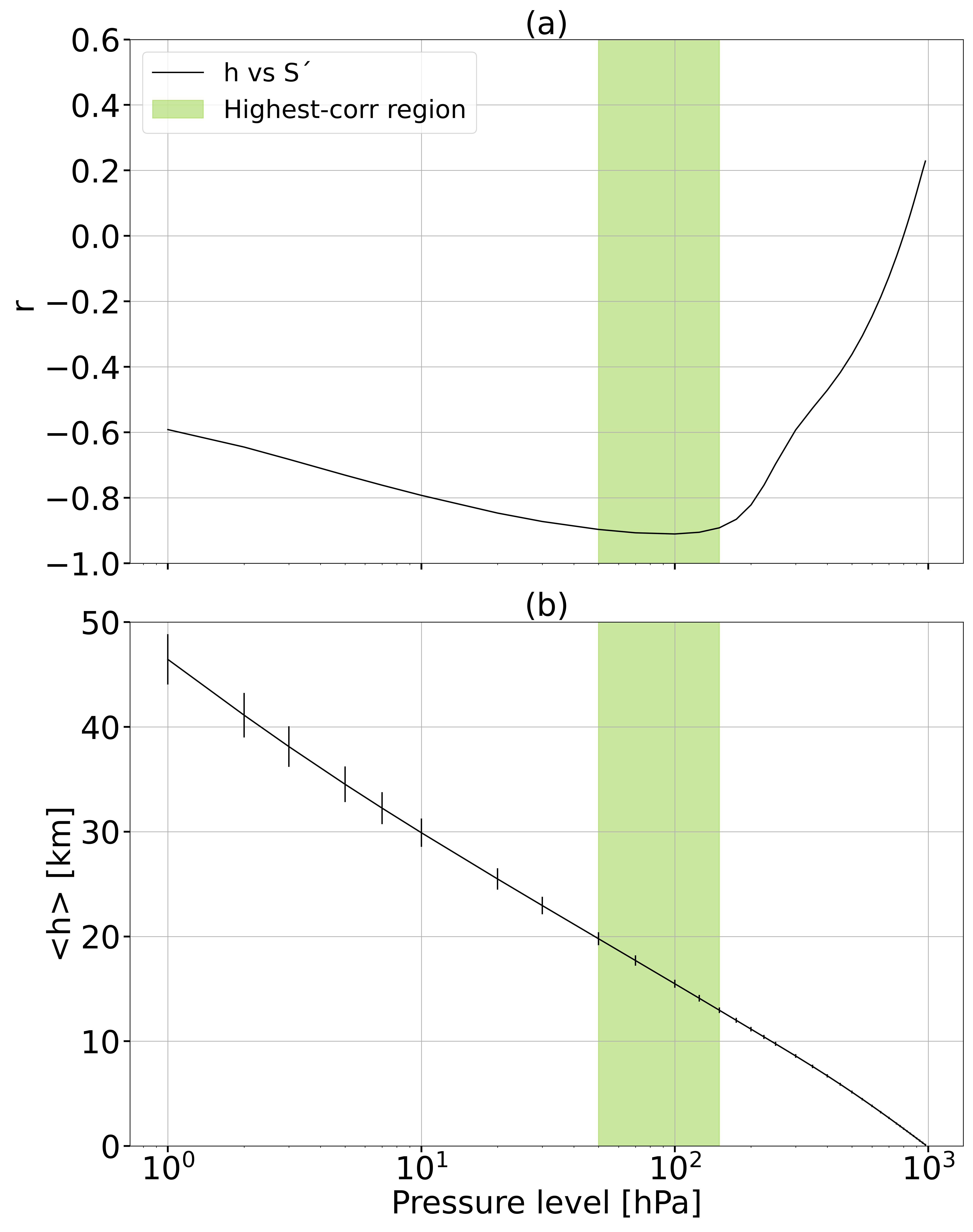}
    \caption{(a) The correlation coefficient $r$ between the pressure corrected count rate ($S'$) and the altitude for each pressure level. The green area represents the region with the highest correlation. (b) The mean altitude $<h>$ associated with each pressure level. The error bars indicate one standard deviation. 
    In both cases the period April 2020-March 2021 is considered.
     }
    \label{fig:fig6}
\end{figure}

Particle showers development depends on the primary type, its energy, zenith angle and the atmospheric depth $\chi$ vertical profile. This mass distribution is  determined by the temperature profile from \hlcyan{the observation level to the atmosphere's upper boundary}. The most relevant variation in $\chi(h)$ is correlated with \hlcyan{the} seasonal cycle (heating-expansion and cooling-compression).

We use \hlcyan{the ERA5's} atmospheric reanalysis to  get the geopotential ($\phi$) at 37 pre\-ssu\-re levels (from 1 hPa to 1000 hPa) at Marambio with \hl{a one-hour} resolution from April 2020 to March 2021.

ERA5 is the fifth generation \hlcyan{model  of the European Centre for Medium-Range Weather Forecasts (ECMWF) for atmos\-pheric reanalysis of the global climate. It provides} hourly estimates of many environmental variables (\citealp{era}).

We consider the height $h$ of each pressure level (directly proportional to $\chi(h)$ assuming $g$ = const).
In order to get this last one, we consider that geopotential height $z$ can be calculated dividing $\phi$ by the Earth's mean gravitational accele\-ra\-tion. 
Then, the height or geometrical altitude is $h$ = $R_e$ $z$/($R_e$\textminus $z$), where $R_e$ is the Earth's radius. This geometric height is relative to \hlcyan{the mean} sea level, and it is assumed that the Earth is a perfect sphere.

On Figure \ref{fig:fig6}\hlcyan{, (a) the Pearson correlation coefficient ($r$) for  $h$  vs. $S'$ can be seen. The region between 50 and 150 hPa (green area) represents the highest correlation} ($r$ $\leq$ \textminus 0.9). This wide region is expected since contiguous pressure levels are highly correlated with each other. The negative temperature effect is dominating, as it was already said at the beginning of this section (maximum intensity in winter and minimum in summer).
On Figure \ref{fig:fig6} (b), the mean altitude $<h>$ for each pressure level is shown. The green region is associated with an altitude\hlcyan{ of about 13 to 20 km,  suggesting} that  variations in the lower stratosphere region are more determinant than \hlcyan{in} any other \hl{la\-yers}.

In this work, we apply the simplest model using only the height of the level with the highest correlation, which is 100 hPa ($r$ =  \textminus 0.91). Furthermore, this is the level where main muon generation is assumed to take place (\cite{dorman}). 
 
\hlcyan{We model the temperature effect on SCRs' counting variability following Equation} (\ref{eq:eq4}).

\begin{equation}
\begin{split}
    \frac{\Delta S '}{S_0 } \times 100 \% =
    \frac{S'(t) - S_0 }{S_0} \times 100 \% \\ 
    =\alpha (H_{100} (t) - H_{100}^0)
    = \alpha \Delta H_{100}
    \label{eq:eq4}  
\end{split}
\end{equation}

$\Delta S '/S_0$ is the pressure corrected counting relative variation, $\alpha$ is the at\-mos\-pheric expansion temperature coefficient given in \% km$^{-1}$ and  $\Delta H_{100} (t)$ is the altitude variation associated with the isobaric level 100 hPa\hlcyan{ with respect to its mean value ($H_{100}^0$ = 15.5 km). The amplitude of $H_{100}$'s variation is about 1 km.} 

In Figure \ref{fig:fig7} \hlcyan{(a), an anti-correlation between $\Delta S '/S_0$ and  $\Delta H_{100}$ can be observed. The effect of the atmospheric expansion on SCRs' counting was estimated to   $\hat{\alpha}$ =  (\textminus 3.89 $\pm$ 0.02) \% km$^{-1}$ (linear temperature coefficient) according to the Equation} (\ref{eq:eq4}). This coefficient has a value slightly below the reported by the muon telescopes of the GMDN using the same model ($\sim$ \textminus 6\% km$^{-1}$ in the northern hemisphere and $\sim$ \textminus 5\% km$^{-1}$ in the southern hemisphere) (\cite{mendoza2016}). We use $\hat{\alpha}$ to remove the temperature effect considering Equation (\ref{eq:eq5}). After that, the variation of the corrected data for pressure and tempe\-rature effects ($S''$) is $<$2\%.

\begin{equation}
   \frac{\Delta S''}{S_{0}} \times 100 \%  =    \frac{\Delta S'}{S_{0}} \times 100 \%  - \hat{\alpha}\Delta H_{100}
\label{eq:eq5}
\end{equation}

We have \hlcyan{chosen} \hl{a one-year} data range to do this analysis in order to remove the seasonal effect. \hlcyan{As seen in Figure} \ref{fig:fig7} (b), \hlcyan{this model explains not only the seasonal variation, but also shorter-term variations (83$\%$ of the total variance).} 

\begin{figure}[ht!]
    \centering
    \includegraphics[width=8cm, height=9cm]{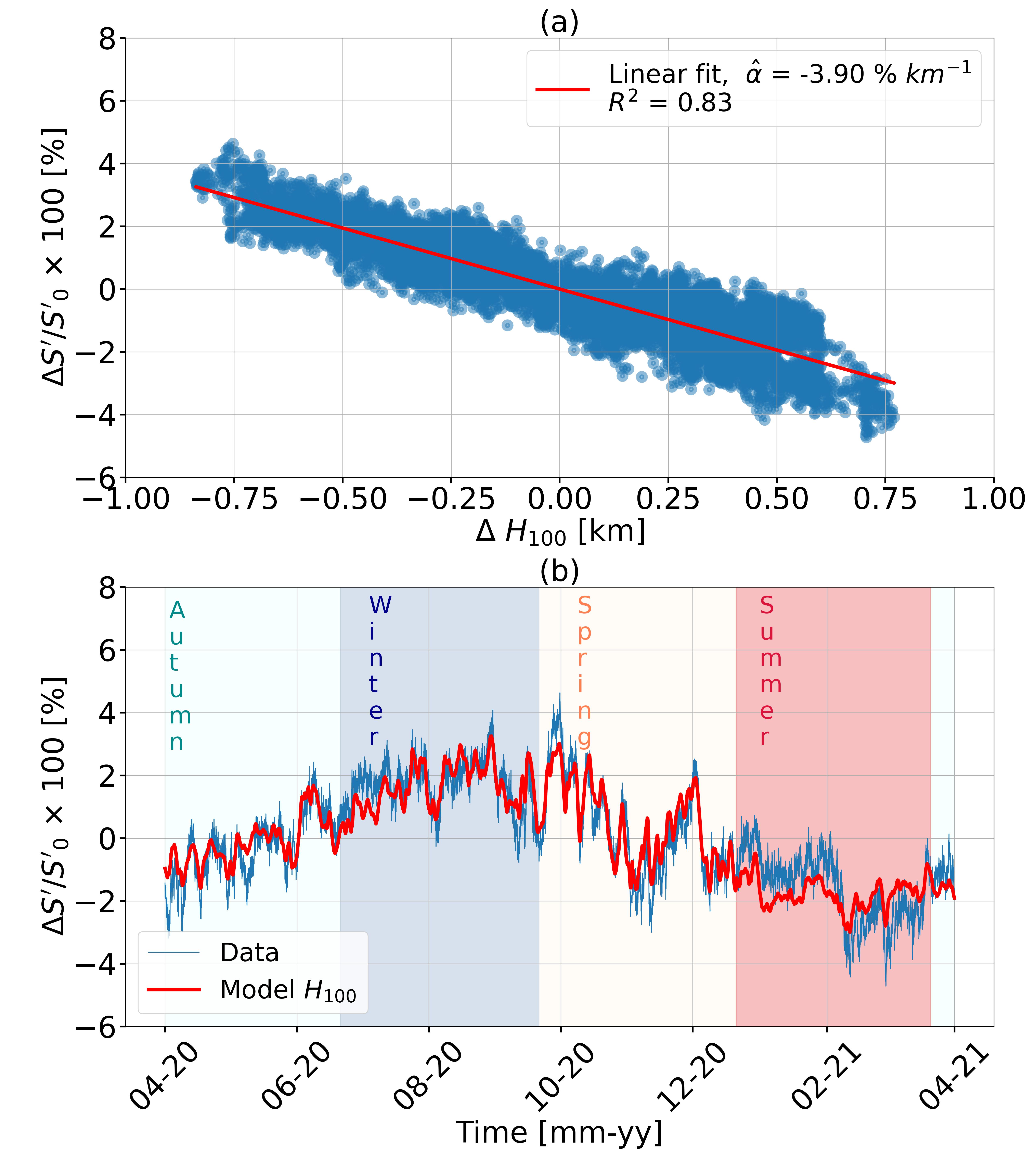}
   \caption{ (a) $\Delta S'/S_0$ $\times$ 100 \% vs. $\Delta H_{100}$ and (b) $\Delta S'/S_0$ $\times$ 100 \% \hlcyan{time series for April 2020-March 2021. 
   The blue points are the data and the red line is a linear model based on Equation} (\ref{eq:eq4}).}
    \label{fig:fig7}
\end{figure}

\section{Daily Variation}
\label{S-mod}

Assuming that after corrections for pressure and temperature there are not more atmospheric variables affecting the count rate, a spectral analysis will provide periodicities related to heliospheric effects.

There are several techniques \hlcyan{to perform a power spectral density (PDS) estimation. Among the nonparametric methods, the  Blackman-Tukey method can be found} (\citealp{BYT}).
This method is based on the fact that the Fourier transform of an autocorrelation function  of a time series is equivalent
to its power spectrum.
Weighting the autocorrelation function by various shapes is a traditional approach to reduce a power leakage.

\begin{figure}[ht!]
    \centering\includegraphics[width=8cm, height=5cm]{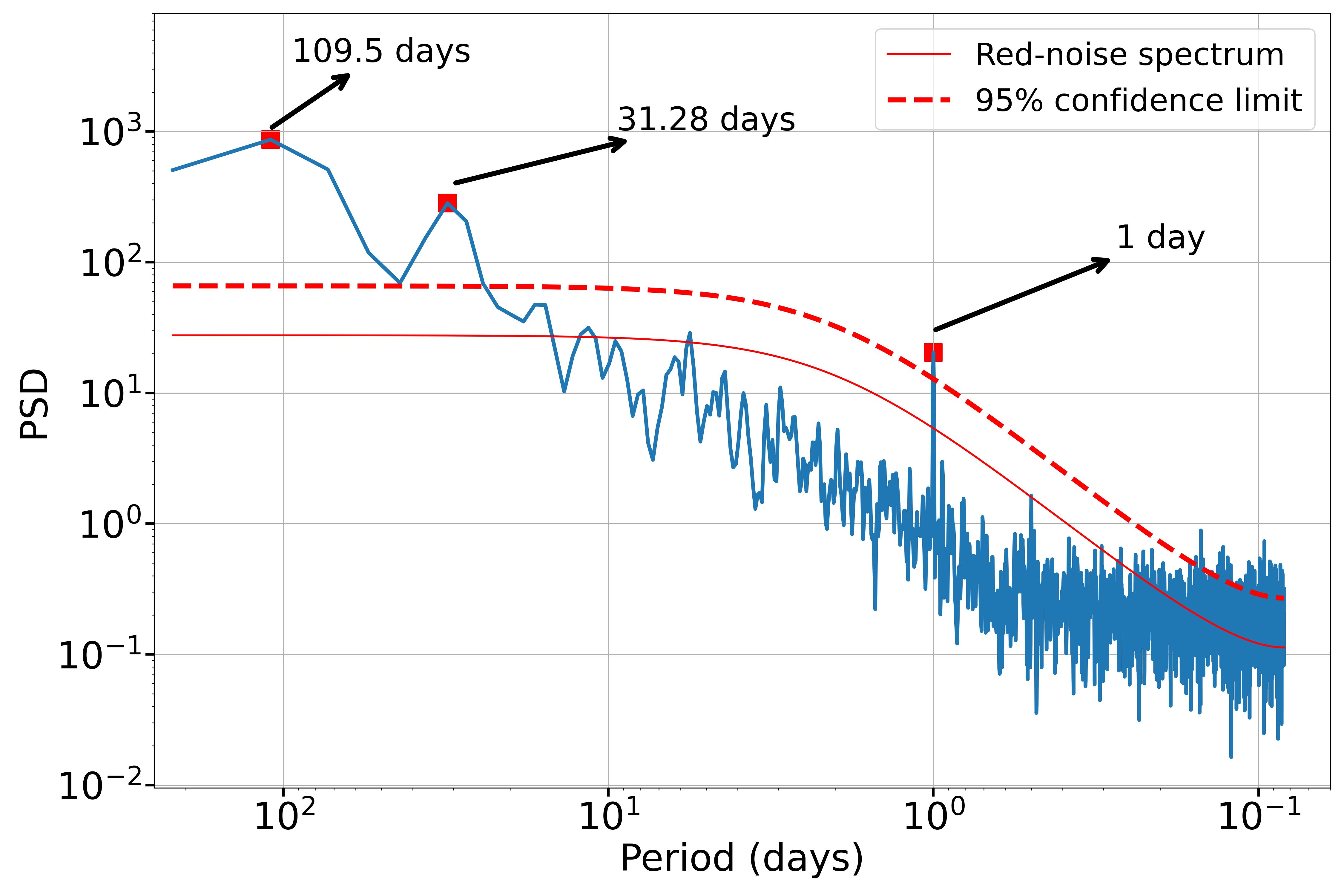}
    \caption{Power spectrum
of the data corrected for pressure and temperature. \hlcyan{The red line represents the red-noise spectrum while the dashed red line marks} the
95 \% confidence limit. Significant periodicities are indicated by red squares.}
    \label{fig:fig8}
\end{figure}

We estimate the PSD of the hourly $S''$ by applying the Blackman-Tukey method and considering the Hanning window of a size equal to 0.3 of the data length.  In order to determine the spectral peak significance, the generation of a red-noise spectrum has been employed. Spectral peaks which have amplitudes of above the 95 \% confidence interval are statistically distinct from the background red-noise spectrum. On Figure \ref{fig:fig8}, \hl{the power spectrum is shown in blue. The red line is the red-noise spectrum, and the dashed red line is the 95 \% confidence limit. The significant periodicities, indicated by red squares, are: 1, 31.28 and 109.5 days.}
These periodicities could be associated with the DV, the solar rotation, and the Rieger periodicity, respectively.

In this work, we focus on the DV, and we used the superposed epoch analysis (SEA) to obtain the typical profiles of this periodicity. The main aim of \hlcyan{the} SEA technique is getting an average profile by taking a sample of individual profiles (\hlcyan{one-day} count rate $S''$ in this case). Each individual profile  has the same number of points or bins in time. The data $S''$ within each bin are averaged to a single value per bin. In our case, we have $N$ = 365 profiles that correspond to 365 days, and each of them has the same bins in time (24 bins of 1 hour long).

\begin{figure}[ht!]
    \centering
    \includegraphics[width=8cm, height=9cm]{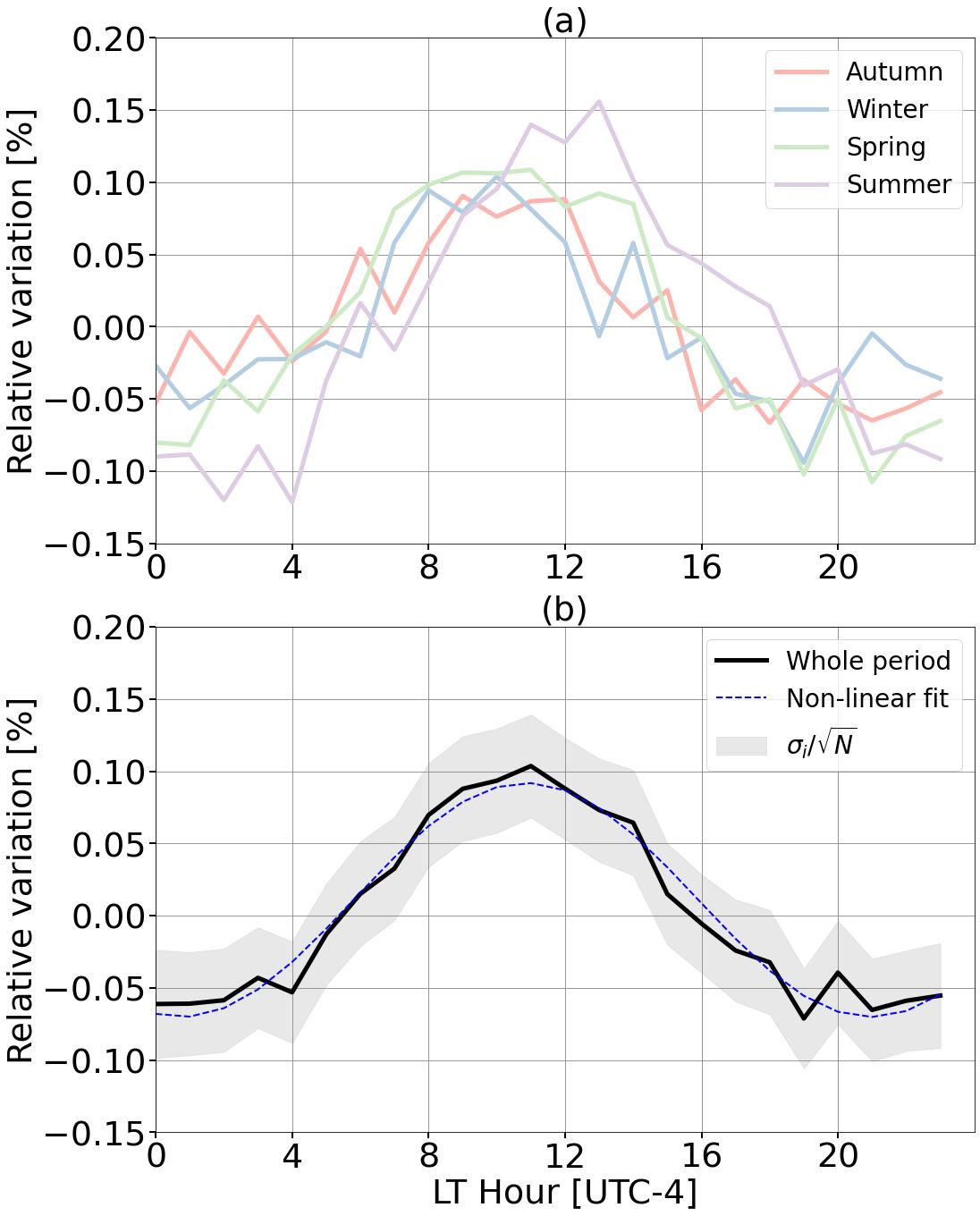}
    \caption{SEA analysis of the temperature and pressure corrected data ($S''$) in each season (a) in the whole period (b).}
    \label{fig:fig9}
\end{figure}

In Figure \ref{fig:fig9} (a) we can see the superposed  profile (relative variation) in each season of the year. In all cases, the maximum value is reached between 8 and 12 hr LT and the amplitude is found between 0.08-0.15 \%. \hlcyan{We expect that}, if there is any residual effect of the expansion and compression of the atmosphere during the year, the profile would look different in each season in terms of phase and amplitude. Since this is not observed, our interpretation is that this modulation has a helios\-pheric and not \hlcyan{an} atmospheric origin.

\hlcyan{The} black curve of Figure \ref{fig:fig9} (b) shows the superposed profile considering \hlcyan{the} entire analyzed year (April 2020 - March 2021). 

The grey region represents the error of the mean of each hourly bin and the blue line is a non-linear fitting based on Equation (\ref{eq:eq7}), assuming a harmonic behaviour as a simplification.

\begin{equation}
    Acos\left(\frac{2\pi}{T}t + \gamma\right) + C
    \label{eq:eq7}
\end{equation}

$T$ = 20.2 hr is the time period, $\gamma$ = \textminus 0.2 rad is the initial phase, $C$ = 0.01 \% is the mean value of the series, and $A$ = \textminus 0.08 \% is the amplitude. The maximum is at 11 hr LT. 

\begin{figure}[ht!]
    \centering
    \includegraphics[width=8cm, height=8cm]{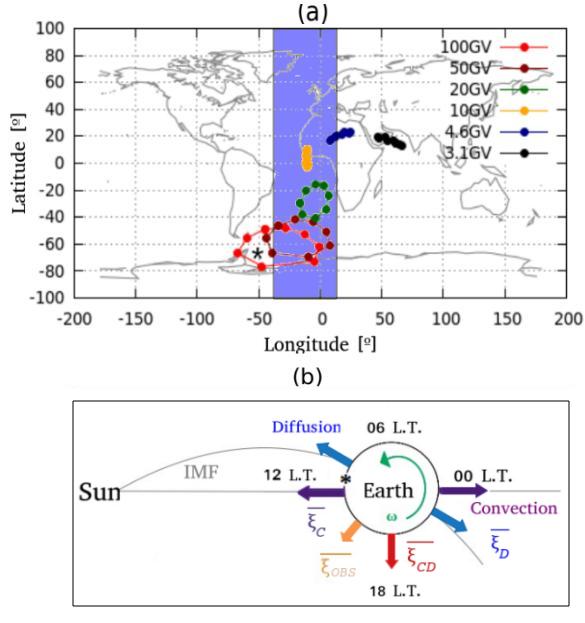}
    \caption{(a) Asymptotic directions of primary protons before interacting with the geomagnetic field, for different energies and different incidence azimuth angles (MAGCOS numerical simulations) adapted from \citealp{Jimmy14}. PCRs that contribute the most to the count observed by Neurus enter through the cones found in the violet region. (b) Diagram of the diurnal variation and the different processes involved: the convective anisotropy ($\bm{\xi_{C}}$), diffusive anisotropy ($\bm{\xi_{D}}$), the predicted anisotropy $\bm{\xi_{CD}}$ by the convective-diffusion theory, and the observed anisotropy $\bm{\xi_{obs}}$. The ``*" indicates Marambio's location.}
    \label{fig:fig10}
\end{figure}

\hlcyan{The DV arises from the spatial anisotropy of the GCR flux reaching Earth due to different physical processes controlling their transport in the interplanetary space. The ground-based detectors record the GCR flux as a rotating lighthouse}, pointing their asymptotic cone to a given direction once a day. As shown in Figure \ref{fig:fig10} (b), the convective-diffusion model assumes two main anisotropies in the ecliptic plane, both point to the direction where the maximum GCR flux comes from: $\bm{\xi_{D}}$ 
a vector parallel/anti-parallel to the IMF and with \hlcyan{an} outward direction related to the \hlcyan{GCRs'} diffusion along IMF; $\bm{\xi_{C}}$ a vector with \hlcyan{an} inward radial direction related to the \hlcyan{GCRs' outward convection by the solar wind}. Assuming that under steady equilibrium conditions there is no net flux of GCRs inward/outward from the Sun, the net spatial anisotropy $\bm{\xi_{CD}}$ is expected to be \hlcyan{found} in the 18 hr LT direction as can be seen in Figure \ref{fig:fig10} (b) (\citealp{forman}).

In order to analyze the observed DV of Figure \ref{fig:fig9} it is ne\-cessary to take into account the asymptotic directions of PCRs arriving \hlcyan{at} Marambio in order to link local time with the spatial IM anisotropy.  An observatory located at a longitude LO records
SCRs generated by PCRs that entered the magnetos\-phere at LO + $\delta$, where $\delta$ is the longitudinal distance between the mean asymptotic cone and the observatory location (LO). Therefore, the direction of the maximum flux in the interplane\-tary space is $\tau$ = LTO + $\Delta$, where LTO is the time of the observed maximum expressed in the observatory LT hour and $\Delta$ = $\delta$ $\times$ 1hr/15$^{\circ}$ is the time difference between both locations.

Based on Figure \ref{fig:fig9} (b), the time of the maximum flux at Marambio is LTO = 11 hr LT (indicated with \hlcyan{an ``*"} in Figure  \ref{fig:fig10} (b)).
Figure \ref{fig:fig10} (a) shows the asymptotic directions for primary protons projected on the Earth’s surface (with rigidities of 3.1, 4.6, 10, 20, 50 and 100 GV) arriving at Marambio (LO = 56$^\circ$) for 15$^\circ$ zenith incidence and eight equispaced incidence azimuth values (figure adapted from \citet{Jimmy14}). Considering that the maximum contribution of PCRs in the counting occurs for energy between 10-40 GeV, Marambio’s mean position of the asymptotic cone is $\delta$ $\sim$ +60$^\circ$ and $\Delta$ $\sim$ +4 hr. 
Finally, the direction of the observed net anisotropy $\bm{\xi_{obs}}$ is $\tau$ = 15 hr LT in interplanetary space with an amplitude of $\sim$ 0.08 \%. This amplitude is smaller than the observed by NMs ($\sim$ 0.3-0.6\%, \hlcyan{depending on the site's latitude) and similar to MTs}. As mentioned in Section~\ref{S-new}, \hl{WCDs are more sensitive to higher primary energies ($\sim$ tens of GeV) than neutron monitors ($<$ 20 GeV). As the solar modulation for primary energies exceeding several tens of GeV is expected to be weaker then, a smaller diurnal amplitude for this type of detector is expected}. In order to have more accurate results of the diurnal anisotropy features, the asymptotic cones of viewing have been taken into account in detail as well as the average PCR energy observed. 

\section{Summary and Conclusions} 
      \label{S-Conclusion}  
\hl{We presented a new CR detector based on the water-Cherenkov effect, which is mainly for space weather purposes. It was installed at the Argentine Antarctic base, Marambio, in 2019.
This detector is part of the Latin American Giant Observatory (LAGO), an extended cosmic rays observatory operating in Latin America and Antarctica.} 

This detector has been measuring continuously mainly particles from EM and muonic components of the particle showers since April 2020. It is necessary to have an accurate estimation of the atmospheric modulations from the data to make it sui\-table for studying solar and interplanetary phenomena \hlcyan{and convenient for SW applications so as to observe} Forbush decreases \hl{and strong ground-level} enhancements. 

In this work, we describe and correct the atmospheric pre\-ssure and temperature effects on the observed SCR flux. We have analyzed observations made during a year (April 2020-March 2021) \hlcyan{during which the laboratory room temperature was at} an optimal range (22 - 24$^\circ$C).
First, we found the well-known anti-correlation between the count rate and the barome\-tric pressure (measured in the laboratory). We have estimated the barometric coefficient per month by performing linear fits. We got  $<$$\hat{\beta}$$>$ =  (\textminus 0.19 $\pm$ 0.02)  \% hPa$^{- 1}$ for the whole period. After removing this effect using $<$$\hat{\beta}$$>$, a clear seasonal effect is observed as a sinusoidal variation with a maximum during winter, a minimum during summer and an amplitude of $\approx$ 3\%.

For the temperature effect, we've worked with data from \hlcyan{ERA5's} atmospheric reanalysis. We observed that
the highest correlation coefficient  between pressure corrected data and altitude occurs in the pressure level corresponding to 100 hPa (mean altitude $H^{100}_0$ = 15.5 km).
The negative temperature effect is dominating. Then we considered the method of effective generation level, which takes into
account altitude variations of this pressure level, which is the muon main production layer. 
We performed a linear fit and got the atmospheric expansion temperature coefficient $\hat{\alpha}$ =  (\textminus 3.89 $\pm$ 0.02) \% km$^{-1}$. This model explained the  $\sim$ 82$\%$ of the total variance.

Finally, we performed a spectral analysis to the corrected data for both effects (barometric and temperature)
and we observed a \hlcyan{significant} periodicity of one day. We analyzed this periodicity using SEA 
and concluded that there are not signifi\-cant changes in the amplitude or phase during different seasons, thus, this periodicity is due to heliospheric phenomena. 
For the whole period, we got an amplitude of $\sim$0.08$\%$ and the time of the maximum is 15 hr LT in the IM, taking into account the effect of the geomagnetic field, which is consistent with the convection-diffusion theory and previous reports. Thus, we conclude that this detector is able to observe spatial anisotropies of \hlcyan{GCR flux as well as other GCRs' properties} of interest for Space Weather studies. 

\section{Acknowledgments}

N.A.S. is fellow of CONICET. S.D. and A.M.G are members of the Carrera del Investigador Científico, CONICET.

This work was supported by the Argentinean grants PICT 2019-02754 (FONCyT\--ANPCyT) and UBACyT\--20020190100247BA (UBA). LAGO collaboration is very grateful to all the participating
institutions and The Pierre Auger collaboration for their continuous support.

\bibliographystyle{jasr-model5-names}
\biboptions{authoryear}
\bibliography{mymanuscript}

\end{document}